\documentclass[12pt]{article}

\usepackage{amsfonts}
\usepackage{graphicx}
\usepackage{amsmath}
\usepackage{natbib}

\begin{document}


\begin{titlepage}

\begin{center}
\bf \large {Correlation Functions of the Magnetization in Thin Films}
\end{center}

\vskip 1.0cm

\begin{center}

{A. Rebei$^{1}$, M. Simionato$^{2}$, G. J. Parker$^{1}$ \\
$^{1}$Seagate Research Center, Pittsburgh, PA 15222 \\
$^{2}$Department of Physics and Astronomy, University of Pittsburgh, PA 
15260\\
}

\end{center}

\vskip 2.0cm


\begin{abstract}
We calculate the correlation functions of uniform magnetization
in thin ferromagnetic films for small deviations from equilibrium, by using
a functional formalism. To take account of dissipation
and fluctuations consistently, the magnetization is coupled to
a bosonic heat bath.  The correlation functions show strong dependence 
on the nature of the coupling between the bath and the system. \ 
Depending on what coupling we choose, we show how the recent  
results $\left( \text{J. Appl. Phys.} 90, 
 5768 (2001); \;  \text{Phys. Rev. B }65, \right. \\ 
\left.  172417 (2002) \right)$ obtained by macroscopic methods 
can be related to the microscopic treatment adopted here.
\end{abstract}


\end{titlepage}


\section{Introduction}

\bigskip The problem of magnetic noise in 
nano-systems and in particular in giant-magnetoresistive 
(GMR) heads  is of considerable
importance to the physics of 
magnetic recording.\ For macroscopic systems, 
the Landau-Lifshitz-Gilbert
equation (LLG) has been used frequently to study thermal 
fluctuations
in magnetic systems mainly through a stochastic 
approach.\cite{brown} 
 \ Magnetic noise in thin anisotropic films has  been 
recently
 treated by at least  a couple 
of different methods. \cite{smith,safonov,jury} The first work 
\cite{smith}  
is based on a linearized LLG and application of the 
fluctuation 
dissipation theorem (FDT).\cite{kubo} \ The second 
method 
\cite{safonov} is also approximate and based  
 on analogies with the 
harmonic oscillator.\cite{lyberatos} \ Both 
calculations are classical, however 
they give different answers. \ Further, in both works it is assumed that 
the system is close to equilibrium.   \ To better 
understand this discrepancy, we study this particular 
problem of magnetic noise in thin films 
 from a microscopic point of view. \ Since we are dealing 
with a 
magnetization slightly disturbed from equilibrium, we 
use 
 bosonic degrees of freedom to describe the 
magnetization.\cite{kittel} \ Moreover, we use the 
language of coherent states (CS) 
to describe the states of the 
magnetization.\cite{glauber}      \ The 
dissipation is simulated by coupling the 
magnetization to a bosonic
environment. \cite{caldeira} \ The noise spectrum 
is found by calculating the 
correlation functions of the magnetization. \ We use a functional
method  borrowed 
from Field Theory to carry out the general 
calculation. \cite{schwinger,
chou} \ These methods are attractive because they are
equally applicable to highly 
non-equilibrium situations and very suitable 
for the many-body problem. \ Two of us have already used 
these methods in a recent paper that 
addressed the conditions under which a LLG equation can 
be recovered from 
a simple quantum model. \cite{rebei} \ The results 
presented here complement 
those presented in Ref. \cite{rebei}. \ However 
this paper can be read 
independently of our previous work. \ The major result 
of this work is a general expression for the correlation functions 
from which  we can recover both LLG-type correlation functions 
and oscillator-like correlation functions. \ Currently experiments
do seem to favor the LLG result however we will not 
address these questions here. \cite{jury,experiment} \ We simply 
show that 
for systems close to equilibrium, we can have different theories 
for a large spin. \ However, it must be kept in mind that the 
results obtained are based on a very simple coupling between 
the magnetization and the bath. \ A more realistic coupling such as
that of conduction electrons interacting with localized magnetic 
moments  
is treated elsewhere.\cite{rebei2} \ In this latter
case  the use of functional 
methods is well justified.   

\bigskip

The paper is organized as follows. In Sec. II, we 
first introduce the Hamiltonian 
for the system considered here. \ Then we write this Hamiltonian
in an approximate form valid for near-equilibrium cases. \ It 
is shown that in this case the spin Hamiltonian is 
exactly that of an oscillator. \ Since correlation functions 
of the damped harmonic oscillator in a thermal bath are available in the 
literature, we deduce from them the corresponding correlation 
functions for the magnetization. \ This result does not apply to 
LLG-type correlation functions, however, which are deduced 
from a different coupling between the spin and the bath. \ In Sec. 
III, we 
review the 
harmonic oscillator CS and their 
relation to those of spin CS.  \ We also write the CS bosonic propagator 
at finite temperature to be used in subsequent sections and 
introduce a generating
functional for these propagators in real-time. \ In 
Sec. IV, we 
derive the CS generating functional for the system 
considered here. \ In Sec. V
, a normal mode analysis of the noise is carried out. \ Here 
we recover 
the Safonov-Bertam \cite{safonov} result in the limit when the 
range of frequencies 
are around the resonance (FMR) frequency. \ In Sec. VI, 
we derive the
general correlation functions for the magnetization. \ The 
LLG result 
   is  shown to follow from the general 
result by assuming a bath for which the product of the 
density of states and the coupling constants is linear 
with frequency. \ This is 
the same condition recovered in Ref. \cite{rebei} and 
is independent of the Hamiltonian of the spin system alone. In 
Sec. VII, we summarize our results. \ Finally in an appendix 
we show how these results can also be obtained form methods 
of equilibrium statistical mechanics.

{\section {Quantum Model} }

In this section, we introduce the quantum model and 
approximate the magnetization operator by a Bosonic operator 
stressing the analogies with the harmonic oscillator. \ This 
picture is in general true for close to equilibrium 
states, irrespective of the equation of motion we are using, 
i.e., LLG or others.

 We consider the following system:  A thin
magnetic slab with easy axis along the z-axis and a hard axis 
along the x-axis which 
are in-plane.  \ We assume that
there is a large external magnetic field $\mathbf{H}$ along the easy 
axis that keeps the average magnetization in-plane. \ We will
be interested only in fluctuations around the equilibrium 
position of the
magnetization, i.e., fluctuations in the x and y 
components of the
magnetization. \ The Hamiltonian for such a system has the 
general form  $\left( \hbar = 1 
\right)$

\begin{eqnarray}
\widehat{\mathcal{H}}  & = & -H \widehat{S}_{z}-K_{1} \widehat{S}_{z}^{2} 
+ 
K_{2}\widehat{S}_{x}^{2}  +\sum_{k}\omega_{k}b_{k}^{+}b_{k} + V \left( 
b_{k},
b_{k}^{\dag}, \widehat{\mathbf{S}} \right),  \label{Ham} 
\end{eqnarray}
where  \ $K_{1}$ and 
$K_{2}$ are the anisotropy constants. \ $\omega_{k}$ 
is the energy 
of the $k$-th bath's oscillator and $\gamma_{k}$'s are the 
coupling constants, 
which can be time-dependent. The spin-bath 
interaction $V$ will be taken linear in 
$\widehat{\mathbf{S}}$ and the bath variables $b_{k}$. A possible 
form for $V$ is
\begin{equation}
V \left( b_{k},
b_{k}^{\dag}, \widehat{\mathbf{S}} \right)=
\sum_{k}\left(  \gamma_{k}^{\ast} \widehat{S}_{+}b_{k}
+\gamma_{k}b_{k}^{+}\widehat{S}_{-}\right).  
\end{equation}
The Bosonic-type 
operators for the bath oscillators, $b_{k}$ satisfy the usual commutation
relations,

\begin{equation}
\left[  b_{k},b_{k}^{+}\right]  =\delta_{kk^{\prime}}%
\end{equation}
For the magnetization, $\widehat{\mathbf{S}}$,  we have the 
following commutation relation

\begin{equation}
\left[  \widehat{S}_{+},\widehat{S}_{-}\right]  = 2 \widehat{S}_{z},
\end{equation}
where
\begin{eqnarray}
\widehat{S}_{+}= \widehat{S}_{x} + i \widehat{S}_{y} ,\\
\widehat{S}_{-}= \widehat{S}_{x} - i \widehat{S}_{y}.
\end{eqnarray}
 \ $K_{1}$ and 
$K_{2}$ are the anisotropy constants. \ $\omega_{k}$ 
is the energy 
of the $k$-th oscillator and $\gamma_{k}$'s are the 
coupling constants, 
which can be time-dependent. The Hamiltonian is therefore comprised of 
three
terms: the magnetization (spin term), the spin-reservoir interaction
 and the reservoir,
\begin{equation}
\widehat{\mathcal{H}}= 
\widehat{\mathcal{H}}_{S}+\widehat{\mathcal{H}}_{SR}+
\widehat{\mathcal{H}}_{R}.
\end{equation} 

If $S$ is the magnitude of the magnetization, we are interested in 
the case in which

\begin{equation}
\frac{\left|  \widehat{S}-\widehat{S}_{z}\right|  }{2S}<<1 ,
\end{equation}
that is deviations from the z-axis are small. 
For precession around the equilibrium 
position, i.e., the z-axis, $S_z$ can be assumed constant and
the Hamiltonian expression for the spin part can be simplified to be 
of the general form
 
\begin{equation}
\widehat{\mathcal{H}}_{S} = \frac{1}{2}\left(  A \; \widehat{S}_{x}^{2} 
+ B \; 
\widehat{S}_{y}^{2}\right),
\end{equation}
where 
\begin{equation}
A=K_1+K_2,\quad B=K_1\;.
\end{equation}
On average, we have $\langle S_{x}\left( t \right) \rangle = 
\langle S_{y}\left( t \right) \rangle = 0$.\ 
To account for thermal fluctuations, we need to 
calculate the two-point correlation
functions of the components of the magnetization, i.e.,
\begin{equation}
\left\langle \widehat{S}_{x}\left(  t\right)  \widehat{S}_{x}
\left(  0\right)  \right\rangle
,\left\langle \widehat{S}_{y}\left(  t\right)  \widehat{S}_{x}
\left(  0\right)  \right\rangle
\quad \ldots
\end{equation}
These are the principal quantities that are needed 
for a full account of 
the noise or dissipation of the magnetization 
  problem for the model considered here. \ Since 
the Hamiltonian is 
quadratic, all higher correlation functions are zero.

It is now more appropriate to  define the following 
operators, \cite{kittel}

\begin{equation}
a=\frac{1}{\left(  2S\right)  ^{\frac{1}{2}}} \widehat{S}_{+}%
\end{equation}
and
\begin{equation}
a^{+}=\frac{1}{\left(  2S\right)  ^{\frac{1}{2}}} \widehat{S}_{-},
\end{equation}
then, we have
\begin{equation}
\widehat{S}_{z}\approx S-a^{+}a
\end{equation}
If we normalize by $2S,$ we simply have
\begin{eqnarray}
a= \widehat{S}_{+},\\
a^{+}= \widehat{S}_{-}
\end{eqnarray}
and
\begin{eqnarray}
\widehat{S}_{x}=\frac{1}{2}\left(  a+a^{+}\right), \\
\widehat{S}_{y}=\frac{1}{2i}\left(  a-a^{+}\right).
\end{eqnarray}
Therefore
\begin{equation}
\left|  \frac{1}{2}-S_{z}\right|  <<1 \; .
\end{equation}

The operators $a$ and $a^{+}$ then behave 
as bosonic degrees of freedom, i.e.,
the magnetization behaves, in this approximation, like a
 harmonic oscillator, \cite{vanvleck}
\begin{equation}
\left[  a,a^{+}\right] =\frac{\widehat{S}_{z}}{S}\approx1 .
\end{equation}
If we rewrite the Hamiltonian in terms of 
these operators, we find 
\begin{eqnarray}
\widehat{\mathcal{H}}  & = & \Omega a^{+}a + 
V \left(  a, a^{+} \right) \\
&&  +\sum_{k}\omega_{k}b_{k}^{+}b_{k}-\sum_{k}\gamma_{k}\left(  
a^{+}b_{k}
+b_{k}^{+}a\right) ,\nonumber
\end{eqnarray}
where  
\begin{equation}
\Omega=H+K_{1}+\frac{1}{2}K_{2},
\end{equation}
and the potential $V$ is in this case equal to
\begin{eqnarray}
V\left(  a^{+},a\right)  =
\frac{1}{4}K_{2}\left(  aa+a^{+}a^{+}\right). \label{potential}
\end{eqnarray}
In general,  the spin part has the form
\begin{eqnarray}
\widehat{\mathcal{H}}_{S}=\Omega a^{+}a+V\left(  a^{+},a\right).
\end{eqnarray}
The calculation presented below can be tailored to deal 
with any polynomial
$V$ which will give rise to some type of 
Feynman rules.\cite{orland}

\bigskip

Before we end this section, we would like to point out 
 the analogy 
between
the spin problem and the harmonic oscillator problem when they 
are coupled to a bosonic bath.  \ For a quantum 
oscillator, the Hamiltonian is 
\begin{equation}
\widehat{\mathcal{H}}_{0  }=\frac{\widehat{p}^{2}}{2M}+
V\left(  \widehat{q}\right)  =\frac{\widehat{p}^{2}}{2M}+\frac{1}%
{2}M\omega_{0}^{2}\widehat{q}^{2},
\end{equation}
while for the spin Hamiltonian, we have
\begin{eqnarray}
\widehat{\mathcal{H}}_{S} & = & \frac{A}{2}\widehat{S}_{x}^{2} +
\frac{B}{2} \widehat{S}_{y}^{2}  \\
& = & \frac{\widehat{S}_{y}^{2}}{2M}+
\frac{1}{2}M\omega_{0}^{2}\widehat{S}_{x}^{2} 
\end{eqnarray}
which means
\begin{equation}
A = M\omega_{0}^{2},\qquad B=\frac{1}{M}. \label{analogy}
\end{equation}
The commutation relation for the harmonic oscillator 
\begin{equation}
\left[  \widehat{q}, \widehat{p}\right]  =i ,
\end{equation}
and for a spin, we similarly have
\begin{equation}
\left[   \widehat{S}_{x},  \widehat{S}_{y}
\right]  =i  \widehat{S}_{z}.
\end{equation}
If $\widehat{S}_{z}$ is a constant of motion, we can replace the operator 
$\widehat{S}_{z}$ by its average value and normalize the remaining 
components by it. \  
Both systems, the harmonic oscillator and the spin,  
are equivalent if they are coupled the {\it{same}} way to the 
bath. \ Most works on the harmonic oscillator case 
involved linear coupling to the bath. Reference 
\cite{grabert} gives an exhaustive treatment of this 
problem. \ If for each $b_{k}, b_{k}^{\dag}$, we define two 
new real operators $x_{k}$ and $p_{k}$, such that 
\begin{equation}
b_{k} = \frac{1}{\sqrt{2}} \left( \sqrt{\omega_{k}} 
x_{k} + i \frac{p_{k}}{\sqrt{\omega_{k}}} \right)
\end{equation}
and
\begin{equation}
b_{k}^{+} = \frac{1}{\sqrt{2}} \left( \sqrt{\omega_{k}} 
x_{k} - i \frac{p_{k}}{\sqrt{\omega_{k}}} \right).
\end{equation}
Then the coupling is given by
\begin{equation}
\widehat{\mathcal{H}}= - q \sum_{k=1}^{N} \gamma_{k} x_{k} \label{coupling}
\end{equation}

\noindent The equations of motion are given by the Heisenberg 
equation. \ For the harmonic oscillator, we have
\begin{eqnarray}
 i \overset{\cdot}{  \widehat{q}}=\left[   \widehat{q}, \widehat{H} 
\right],
\;\;\;\; \;  i \overset{\cdot}{ \widehat{p}}=\left[   \widehat{p}, 
\widehat{H} \right],
\end{eqnarray}
with similar equations for the magnetization with $q \rightarrow
\widehat{S}_{x}$ and $p \rightarrow \widehat{S}_{y}$.
For the linear coupling, Eq.$\left( \ref{coupling} \right)$,
the correlation function of the position,
\begin{equation}
C_{qq}\left( t \right )= \frac{1}{2}\langle q\left( t \right )q
+ q  q\left( t \right ) \rangle ,
\end{equation}
is equal to \cite{grabert}
\begin{equation}
C_{qq}\left( t \right )=  \frac{1}{M} \int \frac{d \omega}{2 \pi} 
\frac{ \omega \alpha \left( -i \omega \right)}{ \left(
\omega_{0}^{2} - \omega^{2} \right)^{2} + \omega^{2} \alpha 
\left( -i \omega \right)^{2}} \; \coth\frac{\beta\omega}2\cos \left( 
\omega t \right),
\end{equation}
where 
\begin{equation}
\alpha \left( z \right) = \frac{1}{M} \sum_{k=1}^{N} 
\frac{\gamma_{k}^{2}}{2 \omega_{k}} \; \frac{2 z}{\omega_{k}^{2} + z^{2}},
\end{equation}
and $\beta$ is inverse temperature $1/kT$ with $k$ the Boltzmann constant. 
\
For Ohmic dissipation, the damping kernel is without 
memory and hence 
$\alpha \left( z \right)$ is a constant. \ This is achieved 
by having a bath with spectral density linear in frequency. \ In 
this 
case the magnetization 
correlation function for the $x-$component in 
the high temperature limit is 
\begin{equation}
C_{xx} \left( \omega \right) = 2 \alpha k T \;  \frac{B}{ \left(
\omega_{0}^{2} - \omega^{2} \right)^{2} + \left(
 \alpha \omega \right)^{2}}. \label{harmonic}
\end{equation}
From Eq. $\left( \ref{analogy} \right)$,  the (FMR) 
frequency $\omega_{0}$ is, as expected, equal to 
$\sqrt{A B}$. \ This result may not  
 however be observed for
a spin variable, since the interaction term has the 
unusual form of having only one component coupled to the bath,
\begin{equation}
\widehat{\mathcal{H}}_{SR} = - \sum_{k=1}^{N} \frac{\gamma_{k}}{2\sqrt{2 
\omega_{k}}} \left(
b_{k} + b_{k}^{+} \right) \widehat{S}_{x}.
\end{equation} 
Here it is the $x-$component of the magnetization 
 which is coupled to the bath. \ Hence the 
 coupling in Eq. 
$\left( \ref{Ham} \right)$ seems more reasonable for 
a spin variable, in general. \ The question of counter-terms 
will not be treated here and any shift in the frequency 
will be absorbed. \ Reference \cite{ford} gives 
a detailed treatment of these terms within the models treated
here.

\bigskip

{\section {The 
Coherent State Representation : Equilibrium and Non-Equilibrium
Dynamics }}

Here we first review the Bosonic coherent state representation
and then show how it can be used within the path-integral 
formulation of quantum mechanics. \ Some typical 
formulas are presented  here for transition 
rates between two states. \ All the results  
are based on the standard Gaussian formula for path integrals.

{\subsection {Coherent States}}

\ Coherent 
states are the natural representation for semi-classical 
calculations. \ A Gaussian wave-packet for a 
harmonic oscillator with minimum uncertainty is a 
coherent state. 
 \  They are formally defined as 
eigenstates of the annihilation
operator \cite{glauber}

\begin{equation}
a\left|  \alpha\right\rangle =\alpha\left|  \alpha\right\rangle
\end{equation}
where $\alpha$ is a complex number. \ Usually the 
ground state is defined as the state with zero quanta,
\begin{equation}
a\left|  0\right\rangle =0.
\end{equation}
For spin coherent states, the ground state is 
taken to be the state with the
highest weight $J$, or zero deviation from highest weight
\begin{equation}
\left|  0\right\rangle_{S} =\left|  J,J\right\rangle ,\qquad S^{2}= 
J\left(  J+1\right)  .
\end{equation}
For Bosons, the Hilbert space ( or space of all possible states) is a 
linear
combination of all state vectors, $
\left\{  \left|  n\right\rangle \right\} $, 
such that
\begin{equation}
\left|  n\right\rangle =\frac{1}{\sqrt{n!}}\left(  a^{+}\right)  ^{n}\left|
0\right\rangle .
\end{equation}
These states form an orthonormal basis. \ Coherent states 
may
 then be written in terms of these states. \ We have
\begin{equation}
\left|  \alpha\right\rangle =e^{\alpha a^{+}}\left|  0\right\rangle.
\end{equation}
Spin coherent states are defined in a similar way. \ We have
\begin{equation}
\left|  \alpha\right\rangle =\frac{1}{1+\left|  \alpha^{2}\right|  
}e^{\alpha
a^{+}}\left|  0\right\rangle_{S},
\end{equation}
where the extra factor in front is due to the constraint that 
$S^{2}$ is constant. 
An important operator relation
 for a path-integral representation is 
the decomposition of the unit operator in terms of coherent  
projection operators 
\begin{equation}
\int\frac{d\alpha^{\ast}d\alpha}{2\pi i}e^{-\alpha^{\ast}\alpha}\left|  
\alpha
\right\rangle \left\langle \alpha\right|  =\widehat{1},
\end{equation}
which is used in the discretization of the 
path integral. \cite{orland}
The coherent states form an over-complete basis. 

In all the calculations carried out below, we keep repeatedly  
 using the fundamental result for the
Gaussian integral,
\begin{eqnarray}
\lefteqn{ \int\underset{i}{\Pi}d\mu\left(  \phi_{i}\right)  \exp\left\{  
-\sum_{ij}%
\phi_{i}^{\ast}A_{ij}\phi_{j}+\sum_{i}\alpha_{i}^{\ast}\phi_{i}\right.
\left.  +\sum_{i}\phi_{i}^{\ast}\alpha_{i}\right\}  }\nonumber \\
& &   \;\;\;\;\;\;\;\;\; = \frac{1}{\det A}\, \,  \,\exp\left\{  
-\sum_{ij}\alpha_{i}^{\ast}\left(
A^{-1}\right)  _{ij}\alpha_{j}\right\}. \label{gaussian}
\end{eqnarray}

{\subsection{ Propagators}}

\bigskip

A typical propagator that shows up often in the 
 calculations of the  correlation functions 
is associated with a Hamiltonian that has the 
following general form,

\begin{equation}
\widehat{\mathcal{H}}\left[ {J}, {J}^{\ast}
 \right]  = \Omega a^{+}a-J(t)a^{+}-J^{\ast}(t)a
\end{equation}
where $J(t)$ is a time-dependent 
external source. \ The kernel of
the evolution operator from an initial state $z_{i}$ 
to a final state $z_{f}$     is  given by \cite{orland}
\begin{equation}
\mathcal{K}_{J}\left(  \overline{z}_{f},t_{f},z_{i},t_{i}\right)  
=\left\langle
z_{f},t_{f}\quad|\quad z_{i},t_{i}\right\rangle =e^{iS \left[ 
{J}, {J}^{\ast}
 \right] }.
\end{equation}
Applying the Gaussian formula, Eq.$\left( \ref{gaussian} 
\right)$, with the boundary conditions
\begin{eqnarray}
z \left( t_{i} \right) & = &  z_{i}, \\
\overline{z} \left( t_{f} \right) & = &  \overline{z}_{f},
\end{eqnarray}
gives the phase $S$,
\begin{eqnarray}
 S \left[ {J}, {J}^{\ast}
 \right] & = & -i \overline{z}_{f} e^{-i\Omega\left(  t_{f}-t_{i}\right)  }
+ \int_{t_{i}
}^{t_{f}}dt\left[  \overline{z}_{f}e^{-i\Omega\left(  t_{f}- t\right)
}J(t)  +J^{\ast}(t)e^{-i\Omega\left(  t-t_{i}\right)  }z_{i}\right]  
\label{phase}\\
& & + i \int_{t_{i}}^{t_{f}}dt \; \int_{t_{i}}^{t_{f}}dt^{\prime}
 \; J^{\ast}
\left( t \right) e^{-i\Omega\left(  t - t^{\prime}\right)  }
J \left( t^{\prime} \right) \Theta\left(
t-t^{\prime}\right) \nonumber.
\end{eqnarray}
$\Theta \left( t \right)$ is the unit step function. \ In using coherent 
states, the variables $z$ and $\overline{z}$ are not necessarily related
by complex conjugation. \ They have to be treated as independent.

\bigskip

{\subsection{Partition Function: Euclidean formulation}}

This formulation is useful 
to calculate thermodynamic equilibrium properties
of a system. \ In this case, the partition function is obtained 
through a calculation of diagonal propagators in imaginary 
time, 
\begin{equation}
\tau=i t,  \qquad   0 \le \tau \le \beta
\end{equation}
In the coherent state representation, the partition 
function is written 
\begin{equation}
\mathbb{Z}_{\beta}\left[  J, J^{\ast} \right]  =
\int d\mu(z)e^{-z^{\ast}z}\left\langle z\left|
\mathbf{T} \exp\left\{-\int_{0}^{\beta}d\tau \widehat{\mathcal{H}} 
\left[J, J^{\ast} \right] \right\} \right|  z\right\rangle ,
\end{equation}
where $\mathbf{T}$ is the imaginary time ordering operator.
The integrals in this partition 
function are all Gaussian and hence can be easily calculated 
with the aid of Eq.$\left( \ref{gaussian} \right)$. \ We get
\begin{equation}
\mathbb{Z}_{\beta}\left[  J , J^{\ast} \right]  =  n\left(\omega\right) 
\exp\left\{ 
{\beta\omega} +
  S \left[  J , J^{\ast} \right]\right\}
\end{equation}
where
\begin{equation}
S \left[  J , J^{\ast} \right] 
=\int_{0}^{\beta}d\tau\int_{0}^{\beta}d\tau^{\prime}J^{\ast}(\tau)D\left(
\tau,\tau^{\prime}\right)  J(\tau^{\prime}),
\end{equation}
$D\left(
\tau,\tau^{\prime}\right)$ is the Feynman propagator of 
this model
\begin{equation}
D\left(\tau,\tau^{\prime}\right)= \langle \mathbf{T} 
\left( a^{+}\left(
\tau \right) a \left(\tau^{\prime} \right) \right) \rangle.
\end{equation}
It is not difficult to find that it is given by
\begin{eqnarray}
D\left(  \tau,\tau^{\prime}\right)  = n\left(\omega \right) 
e^{-\omega\left(  \tau
-\tau^{\prime}-\frac{\beta}{2}\right)  }\left[  e^{\beta\frac{\omega}{2}%
}\theta\left(  \tau-\tau^{\prime}\right) 
  +e^{-\beta\frac{\omega}{2}}\theta\left( \tau^{\prime} -\tau \right)
\right],
\end{eqnarray}
where
\begin{equation}
n\left(  \omega\right)  =\frac{1}{e^{\beta\omega}-1}%
\end{equation}
is the Bose-Einstein distribution. \ Its Fourier transform coincides
with the well known Matsubara propagator

$$
D(\omega_n,\omega)=\frac1{i\omega_n+\omega},\quad \omega_n=2\pi \, n\, T,\quad
n=0,\pm 1, \pm 2, ... \;.
$$
\ This last propagator 
is essential for any calculations that involve 
calculating expectation values of any observable.\cite{rivers} \ In 
the appendix we 
show how to use this method to calculate correlation functions.

\bigskip

{\subsection {Real-Time Formulation: Dynamics}}

The real-time formulation deals with non-equilibrium questions. \ In 
this case we can derive equations of motion for any 
observables. \cite{chou} \ This is the method  we  adopt in the 
calculations
of the correlation functions of the magnetization.\ 
For a general operator $\mathcal{O},$ its average value 
at any time $t$
 is given in terms
of the  density matrix $\rho$,
\begin{equation}
\left\langle \mathcal{O} \left( t \right) \right\rangle =Tr\left\langle 
\rho \mathcal{O} \left(t \right) \right\rangle.
\end{equation}
The operator $\mathcal{O}$ is in the Heisenberg picture,
\begin{equation}
\mathcal{O}(t)=e^{i\widehat{\mathcal{H}} t} \mathcal{O} e^{-i 
\widehat{\mathcal{H}} t}.
\end{equation}
Therefore the average of the observable $\mathcal{O}$ at 
time $t$ can 
be written in terms of that at $t=0$,
\begin{equation}
\left\langle \mathcal{O}(t)\right\rangle =Tr\left(  
\rho e^{i\widehat{\mathcal{H}}
t}\mathcal{O}e^{-i \widehat{\mathcal{H}} t}\right).
\end{equation}
This latter average can be written in terms of path integrals
as in the equilibrium case. \ First we define the 
operators $\mathcal{K}$ and $\overline{\mathcal{K}}$. \ The 
operator  $\mathcal{K}$ is a forward propagator and 
is defined as follows,
\begin{eqnarray}
\mathcal{K}\left[  J_{1}, J_{1}^{\ast}    \right]   & = & \mathbf{T} 
\exp \left\{ {-i} \int_{t_{i}}^{t_{f}}\left(
\widehat{\mathcal{H}} -J_{1}^{\ast}a-J_{1}a^{+}\right)  dt\right\}\\
& = & \mathbf{T} \exp\left\{ {-i}\int\left(  
\widehat{\mathcal{H}} -F_{1}^{x}S_{x}-F_{1}^{y}S_{y}\right)
dt \right\}\nonumber
\end{eqnarray}
$\mathbf{T}$ is a time ordering operator. \ $\overline
{\mathcal{K}}$ is a backward operator and is therefore
defined in terms of anti-ordered time operator $\overline{
\mathbf{T}}$,
\begin{eqnarray}
\overline{\mathcal{K}}\left[  J_{2}, J_{2}^{\ast}  
\right]   & = & \overline{\mathbf{T}} \exp\left\{ {-i}
\int_{t_{f}}^{t_{i}}\left(
\widehat{\mathcal{H}}-J_{2}^{\ast}a-J_{2}a^{+}\right)  dt\right\}\\
& = & \overline{\mathbf{T}} \exp \left\{ {-i}\int\left(  
\widehat{\mathcal{H}}-F_{2}^{x}S_{x}-F_{2}^{y}S_{y}\right)
dt \right\} \nonumber,
\end{eqnarray}
where $\mathbf{F}_{1}$ and $\mathbf{F}_{2}$ are real external fields 
which are coupled to the transverse components of 
the magnetization. \ Similar to the 
equilibrium case, we define a generating functional 
\begin{equation}
\mathbb{Z}\left[  \mathbf{J} , \mathbf{J}^{\ast} 
\right]  =Tr\left\{  \rho\left[  J_{3}, 
 J_{3}^{\ast}         \right]  \overline{\mathcal{K}}
\left[
J_{2},  J_{2}^{\ast}  \right]  \mathcal{K}
\left[  J_{1},  J_{1}^{\ast}  \right]  \right\} . \label{gen}
\end{equation}
$\mathbf{J}$ is now the three-vector $\left( J_{1}, J_{2}, 
J_{3} \right)$. \ The density matrix $\rho$ is assumed 
of the form
\begin{eqnarray}
\rho \left[  J_{3}, J_{3}^{\ast}  
\right]   & = & {\mathbf{T}}_{I} \exp \left\{ -
\int_{0}^{\beta}\left(
\widehat{\mathcal{H}}-J_{3}^{\ast}a-J_{3}a^{+}\right)  d\tau \right\} \\
& = & {\mathbf{T}}_{I} \exp \left\{ - \int_{0}^{\beta}\left(  
\widehat{\mathcal{H}} - F_{3}^{x}S_{x}-F_{3}^{y}S_{y}\right)
d\tau\right\} \nonumber,
\end{eqnarray}
where $\mathbf{T}_{I}$ is now a time-ordering operator along the imaginary 
time axis. \
Hence, all correlation functions can be obtained 
from the coefficients of the
Taylor expansion of the functional $ \mathbb{Z} 
\left[  \mathbf{J} , \mathbf{J}^{\ast}  \right]$ around $\mathbf{
J}=\mathbf{J}^{\ast}=0$ (or $\mathbf{F}_{1}=\mathbf{F}_{2} 
=\mathbf{F}_{3} = 0$). \ For example, the average value of the 
$x-$component
of the magnetization at time 
$t$ can be found by differentiating $\mathbb{Z}$ with 
respect to $F_{1}^{x}$ at the same time $t$,
\begin{equation}
\frac{1}{\mathbb{Z}}\left.  \frac{\delta \mathbb{Z}\left[  \mathbf{J}, 
\mathbf{J}^{\ast} \right]  }{\delta F_{1}^{x}
\left( t \right) }\right|  _{F=0}= - \left\langle S_{x}\left( t 
\right) \right\rangle.
\end{equation}
Next we define another functional $\mathbb{W}$ which at equilibrium 
becomes the thermodynamic potential of the system,
\begin{equation}
\mathbb{Z}\left[  \mathbf{F}_{i=1,2,3}\right]  = \exp\left\{ i 
\mathbb{W}\left[  F\right]  \right\}.
\end{equation}
The functional $\mathbb{W}$, as will be seen below, is 
 the more appropriate functional 
to calculate and
expand in powers of $\mathbf{J}$ and $\mathbf{J}^{\ast}$ 
(or $\mathbf{F}$). \ Therefore, we have for averages and two-point 
correlation functions,
\begin{equation}
\left.  \frac{\delta \mathbb{W}}{\delta F_{1}^{x}\left( t
\right) }\right|_{\mathbf{F}=0}=\left\langle
S_{x}(t)\right\rangle
\end{equation}
and
\begin{equation}
\left.  \frac{\delta^{2} \mathbb{W}}{\delta F_{1}^{x}(t)\delta F_{1}^{x}
(t^{\prime}
)}\right|_{\mathbf{F}=0}= - {i}\left\langle \mathbf{T} \left(
S_{x}(t)S_{x}(t^{\prime
}) \right) \right\rangle.
\end{equation}
Similar expressions hold when we differentiate $\mathbb{W}$ with respect
to the sources $\mathbf{J}$ and $\mathbf{J}^{\ast}$. \ They are related 
to each other by chain rule, e.g., we have 
\begin{eqnarray}
\frac{\delta}{\delta F_{1}^{x}}=\frac{\delta}{\delta J_{1}}+\frac{\delta
}{\delta J_{1}^{\ast}}.
\end{eqnarray}
Depending on how we couple the bath to the spin, we use either sources 
to find the corresponding correlation functions. \ In the normal mode 
coupling scheme, we assume that the normal modes of the spin 
are coupled to the normal modes of the bath. \ In this case, it is 
more advantageous to express everything in terms 
of creation and annihilation operators and hence use the $\mathbf{J}$ 
sources
to get the correlation functions. \ This is what we do in section 5. \ In 
section 6, the spin is coupled directly to the bath oscillator. \ Hence 
in this case we use the $\mathbf{F}$ sources to get the 
correlation functions of the spin. \ In the next section, we give 
an explicit expression for the functional $\mathbb{W}$ in terms of 
coherent states.

\bigskip

{\section{Coherent State Generating Functional}}

In this section, we continue working within the real-time formulation. \ 
We give the full expression for the generating functional in the
coherent state representation and calculate all the 
associated propagators in this case.

The generating functional $\mathbb{Z}$ is defined above, Eq.$\left(
\ref{gen} \right)$. \ Using coherent states, for both the bath and spin, 
 this trace 
formula can be written 
in terms of path integrals over spin variables and bath variables,
\begin{eqnarray}
\mathbb{Z}\left[  \mathbf{J}, 
 \mathbf{J}^{\ast}  \right]   & = &\int d\mu(\alpha_{1}) 
d\mu(\alpha_{2})
d\mu(\alpha_{3}) d\mu(\varphi_{1}) d\mu(\varphi_{2})
d\mu(\varphi_{3})  \left\langle \alpha_{1},\varphi_{1}\left|  \rho \left[ 
J_{3}, J_{3}^{\ast} \right] \right|
\alpha_{2},\varphi_{2}\right\rangle \nonumber \\ 
&& \times \left\langle \alpha_{2},\varphi_{2}\left|
\overline{\mathcal{K}}\left[ J_{2}, J_{2}^{\ast} \right] 
\right|  \alpha_{3},\varphi_{3}\right\rangle
\left\langle
\alpha_{3},\varphi_{3}\left|  \mathcal{K} \left[ J_{1}, J_{1}^{\ast} 
\right] \right|  \alpha_{1},\varphi_{1}
\right\rangle   \\
& & \times\exp\left\{  -\left|  \alpha_{1}\right|  ^{2}-\left|  
\alpha
_{2}\right|  ^{2}-\left|  \alpha_{3}\right|  ^{2}\right\}  \exp\left\{
-\sum_{k} \left(  \left|  \varphi_{1,k}\right|  ^{2}+\left|  \varphi
_{2,k}\right|^{2}+\left|  \varphi_{3,k}\right|^{2}\right)  \right\} \nonumber.
\end{eqnarray}

The $\alpha_{i=1,2,3}$ represent states of the spin system, while the
$\mathbf{\varphi}_{i=1,2,3}$ represent the bath states. \ This integral 
can be formally written as a path integral along the path in 
Fig.\ref{path} with periodic boundary conditions similar 
to the equilibrium partition function calculations. \ This functional 
can be calculated exactly only in few cases in particular if 
the Hamiltonian is quadratic. \ Higher order terms can be accounted for 
only approximately. \ This is best done through a graphical 
procedure such as the Feynman diagram technique. \ Here we have a 
quadratic Hamiltonian and hence we can solve for $\mathbb{Z}$, however 
we will mention briefly what happens in the general case.

In our case, the bath degrees of freedom can be integrated out exactly and 
we can derive an exact effective action for the spin degrees of freedom. In
the general case, the effective action can be derived perturbatively.
From it, we will calculate the correlation functions 
of $\mathbf{S}$.\ We find 

\begin{eqnarray}
\mathbb{Z}\left[  \mathbf{J}, \mathbf{J}^{\ast}
 \right]   & = & \int d\mu(\alpha_{1})\int d\mu(\alpha_{2})\int
d\mu(\alpha_{3}) \; \exp\left\{  -\left|  \alpha_{1}\right|  ^{2}-\left|  
\alpha
_{2}\right|  ^{2}-\left|  \alpha_{3}\right|  ^{2}\right\} \\
& & 
\times\int_{\alpha_{1}}^{\overline{\alpha}_{3}}d\mu(z_{1})\int_{\alpha_{3}
}^{\overline{\alpha}_{2}}d\mu(z_{2})\int_{\alpha_{2}}^{\overline{\alpha}_{1
}
}d\mu(z_{3})\;\exp\left\{  \sum_{i=1}^{3}I_{i}\left[  z_{i},
\overline{z}_{i}
,J_{i},J_{i}^{\ast}\right]  \right\} 
 \mathcal{F}\left(  \mathbf{Z},\overline{\mathbf{Z}} \right) \nonumber   
\end{eqnarray}

\noindent where $\mathcal{F}\left( \mathbf{Z},\overline{\mathbf{Z}}
\right)$ is the Feynman-Vernon
functional for the spin-bath system. \ It is given by

\begin{eqnarray}
\ln\mathcal{F}\left( \mathbf{Z},\overline{\mathbf{Z}}  
\right)  =  \int 
dt \int d t^{\prime}  \left[ -
     \sum_{k} \left|
\gamma_{k}\right|^{2}\overline{\mathbf{Z}}\left(  t\right)
\cdot G^{k}\left(  t , t^{\prime} \right) \cdot  \mathbf{Z} \left(  
t^{\prime} \right)  \right],
\end{eqnarray}
where the three-vector $\mathbf{Z}$ is related to the three branches 
of the curve $C$, Fig. \ref{path},
\begin{equation}
\mathbf{Z} =\left(
\begin{array}
[c]{c}%
z_{1}\\
-z_{2}\\
z_{3}%
\end{array}
\right) .
\end{equation}

\noindent The time integrations are defined based on the path $C$:
\begin{eqnarray*}
t_{i}  & <t,t^{\prime}<t_{f}, &  t,t^{\prime} \in C^{(+)},C^{(-)}  \\
t_{i}  & <t,t^{\prime}<t_{i} -i \beta, &   t,t^{\prime} \in C^{(0)}.
\end{eqnarray*}

\noindent The Feynman-Vernon term is the only term which is dependent on 
the 
bath parameters. \ The functions $G^{k}_{ij}\left( t, t^{\prime} \right)$, 
nine in total,  
are propagators associated with the bath oscillators. \ Hence 
they can easily be calculated using Eq.$\left( \ref{phase}  \right)$ since 
the oscillator part of the Hamiltonian is quadratic and the spin can 
be considered as the external field. \ The indices $i,j = 1,2,3$ relate to 
the branches $C^{(+)}, C^{(-)}, C^{(0)}$ of $C$. They are \cite{rivers}

\[
G_{ij}^{k}(t-t^{\prime})=\left[
\begin{array}
[c]{ccc}
  \theta \left(  t-t^{\prime} \right)
G_{21}^{k}
 +\theta\left(  t^{\prime}-t\right)  G_{12}^{k} 
 &  n \left( \omega_{k} 
\right) 
e^{-i\omega_{k}\left(  t-t^{\prime}\right)  }   & 
    \\
   n\left( \omega_{k} \right) 
e^{\beta \omega_{k}}e^{-i\omega_{k}\left(  t-t^{\prime}\right)  }    & 
\theta\left(  
t-t^{\prime}\right)
G_{12}^{k} 
+\theta\left(  t^{\prime}-t\right)  G_{21}^{k}      & 
   \\
 n\left(  \omega_{k}\right)  e^{\beta \omega_{k}}
e^{i\omega_{k}(t+i\tau)}    &  G_{31}^{k} ( t - \tau )   &
  \\
\end{array}
\right.
\]
\begin{equation}
\left.
\begin{array}
[c]{ccc}
& &  n \left(  \omega_{k}\right)e^{-i\omega_{k}(t+i\tau)}   \\
& &   G_{13}^{k}\left(  t,\tau\right)  \\
& &  n \left( \omega_{k}\right)e^{-\omega
_{k}\left(  \tau-\tau^{\prime}-\frac{\beta}{2}\right)  }
 \left[  \theta\left(  \tau-\tau^{\prime}\right)  e^{\omega_{k}\frac{\beta
}{2}}  +\theta\left(  \tau^{\prime}-\tau\right)  e^{-\omega_{k}\frac
{\beta}{2}}\right]
\end{array}
\right].
\end{equation}
\bigskip

\noindent These ``path-coupling'' functions 
show that if the initial time $t_{i}$ is 
taken to be in the infinite past, $t_{i} \longrightarrow -\infty$, the 
branch $C^{(0)}$ decouples from the other two branches. \cite{rivers}
 \ This is the case where transient 
effects have died out. \ In the rest of this paper, these 
transient effects will be neglected and we will concentrate only 
on the real-time paths $C^{(+)}$ and $C^{(-)}$. \ We take account 
of the third branch through the assumption that 
initially the system is in equilibrium.  \ For a general 
 potential $V$, the generating functional can be written in terms of 
that of a free system, $\widehat{\mathcal{H}}_{0}=\Omega a^{\dagger}a$, 
interacting
with the bath,
\begin{equation}
\mathbb{Z}\left[  \mathbf{J}, \overline{\mathbf{J}}  \right]  
=\exp\left\{  - {i}\int_{C}dt \; V\left[
\frac{\partial}{\partial J_{i}(t)},\frac{\partial}{\partial\overline{J}
_{i}(t)}\right]  \right\}  \mathbb{Z}_{SB}\left[  \mathbf{J}, 
 \overline{\mathbf{J}} \;  \right].
\end{equation}
$\mathbb{Z}_{SB}$ is therefore the generating functional 
of a particle interacting with the
bath and there is no external potential. \ This latter 
formula is valid in the general case and is the 
start of any perturbative calculations.  \  The action along the
real-time trajectories is given by
\begin{equation}
iI_{1}^{0}= {i}\int dt\left[  \frac{\overset{\cdot}{\overline{z}%
}_{1}z_{1}-\overline{z}_{1}\overset{\cdot}{z}_{1}}{2i}-\Omega\overline{z}%
_{1}z_{1}\right]
\end{equation}
along the path $C^{\left( + \right)}$ and by
\begin{equation}
iI_{2}^{0}=- i \int dt\left[  \frac{\overset{\cdot}{\overline{z}%
}_{2}z_{2}-\overline{z}_{2}\overset{\cdot}{z}_{2}}{2i}-\Omega \overline{z}%
_{2}z_{2}\right]
\end{equation}
along the path $C^{\left( - \right)}$, Fig. \ref{path}.
At $t_i\rightarrow-\infty$, \  the system is at equilibrium, then
we can assume that the initial density matrix  is thermal,
with $J_{3}(t_i)=0$. 
\ Therefore we write that
\begin{equation}
\rho \left( t_i \right)=\frac1{Z(t_i)} e^{-\beta H(t_i)},\quad 
t_i\to-\infty
\end{equation}
 \ Then, we observe that 
\begin{eqnarray}
\left\langle \alpha_{1}\left|  \rho(-\infty)\right|  
\alpha_{2}\right\rangle
=\int_{\alpha_{1}}^{\overline{\alpha}_{2}}
d\mu(z_{3})e^{iI_{3}^{0}\left[
z_{3},\overline{z}_{3}\right]  }\mathcal{F}\left(  z_{3},
\overline{z}_{3} \right),
\end{eqnarray}
where $I_{3}^{0}$ has the same expression as $I_{1}^{0}$ but with 
$t \rightarrow it$. \ Hence in this case, the initial density 
matrix element is 
just another overall factor in the generating functional 
$\mathbb{Z}$,
\begin{eqnarray}
\mathbb{Z}\left[  \mathbf{J}, \overline{\mathbf{J}} \; 
\right]   & = & \int d\overline{\mu}(\alpha_{3})
 \int d\overline{\mu}(\alpha_{1})d\overline{\mu}(\alpha_{2})\left\langle
\alpha_{1}\left|  \rho(-\infty)\right|  \alpha_{2}\right\rangle \\
& &  \times\int_{\alpha_{1}}^{\overline{\alpha}_{3}}
d\mu(z_{1})\int_{\alpha_{3}
}^{\overline{\alpha}_{2}}\left[  d\mu(z_{2})
  \exp\left[  - \int_{C^{+}} dt \; V\left(  \frac{\partial
}{\partial \mathbf{J}(t)}, \frac{\partial}{\partial \overline{
\mathbf{J}}(t) }
      \right)  \right]  \right] \nonumber\\
& & \times\exp\left\{i I_{1}^{0}\left[  
z_{1},\overline{z}_{1}\right]
+ i I_{2}^{0}\left[ z_{2},\overline{z}_{2}\right]  
 +  i \int dt\left(  \mathbf{J}\cdot\overline{\mathbf{Z}}+
\overline{\mathbf{J}}\cdot
\mathbf{Z}\right)  \right\}\mathcal{F}\left( \mathbf{Z}, 
\overline{\mathbf{Z}} \right), \nonumber
\end{eqnarray}

with $ \mathbf{Z}   =\left(  z_{1},-z_{2}\right)$ and the measure
is defined by
\begin{equation}
d\overline{\mu}\left(  \alpha\right)  =d\mu\left(  \alpha\right)  
e^{-\left|
\alpha\right|  ^{2}}.
\end{equation}
\bigskip Therefore we define a new generating 
functional $\widehat{\mathbb{Z}}$
\begin{equation}
\mathbb{Z}\left[  \mathbf{J},  \overline{\mathbf{J}} \right] = 
\int d\overline{\mu}
(\alpha_{3})\int d\overline{\mu
}(\alpha_{1})
 \left[  \int d\overline{\mu}(\alpha_{2})\left\langle \alpha_{1}\left|
\rho(-\infty)\right|  \alpha_{2}\right\rangle \times
\widehat{\mathbb{Z}}\left[
\mathbf{J}, \overline{\mathbf{J}}\right]  \right] \nonumber .
\end{equation}
We can now adopt a different notation that takes into account 
the path $C$ implicitly by defining a scalar product and 
combine the different components into a single vector. \ The 
generating function becomes
\begin{eqnarray}
\widehat{\mathbb{Z}}\left[ \mathbf{J}, \overline{\mathbf{J}} \; 
\right]  & = & \int 
d\mu(z_{1})\int d\mu(z_{2})\exp\left[  - \int
dt V\left(  \frac{\partial}{\partial \mathbf{J}(t)},  
\frac{\partial}{\partial \overline{\mathbf{J}(t)}}  \right)  \right] 
\nonumber\\
& & \times\exp\left\{  iI^{0}\left(  
\mathbf{Z},\overline{\mathbf{Z}}\right)  
+ i \int dt\left(
\overline{\mathbf{Z}}\cdot \mathbf{J}+\overline{\mathbf{J}}\cdot
\mathbf{Z}\right)  \right\} \mathcal{F}
\left(  \mathbf{Z},\overline{\mathbf{Z}} \right),
\end{eqnarray}
where now the vector $\mathbf{Z}$ is defined 
\begin{equation}
Z=\left(
\begin{array}
[c]{c}
z_{1}\\
z_{2}
\end{array}
\right),
\end{equation}
and the free action is 
\begin{equation}
I^{0}   =\sum_{i=i,j}\sigma^{ij}I_{j}^{0},
\end{equation}
with
\begin{align}
\sigma^{ij}  & =\left(
\begin{array}
[c]{cc}
1 & 0\\
0 & -1
\end{array}
\right).
\end{align}
The complex scalar product is now defined by
\begin{equation}
\overline{\mathbf{Z}}\cdot \mathbf{J}=\sigma^{ij}
\overline{z_{j}} J_{j}.
\end{equation}
This notation makes it possible to take into account 
the closedness of the real-time path by just taking 
one branch of the curve $C$ and  doubling the 
components of the dynamical variables. \ The matrix
$\sigma_{ij}$ plays the role of a metric.

\bigskip Now we turn to some properties satisfied by the functions
$G_{ij}^{k}$. \ These properties are better displayed
in Fourier space. \ The Fourier space representation of
the Feynman propagator is given by
\begin{eqnarray}
G_{11}^{k}\left(  \omega\right)   & = & \int dt \; 
e^{i\omega t}\left( 1 + n\left(
\omega_{k}\right)  \right)  \Theta\left(  t\right)  e^{-i
\omega_{k}t}
 +\int dt \;e^{i\omega t} n\left(  \omega_{k}\right)  \theta\left(  
-t\right)
e^{-i\omega_{k}t}\\
 &  = &  \left( 1 + n\left( \omega_{k}\right) \right)
\frac{i}
{\omega-\omega_{k} + i \epsilon}
 -  n\left(  \omega_{k}\right)  \frac{i}{\omega-\omega_{k}
- i \epsilon},
\end{eqnarray}
where  $\epsilon \longrightarrow 0^{+}$. \ $\mathcal{P}$ 
stands for  the principal part of the integral.
For the anti-time ordered propagator, we have
\begin{eqnarray}
G_{22}^{k}\left(  \omega\right)   & = & \int dt 
e^{i\omega t} n\left(  \omega
_{k}\right)  \Theta\left(  t\right)  e^{-i\omega_{k}t}
 + \int dt e^{i\omega t}\left( 1 + n\left(  \omega_{k}\right) \right)
  \Theta\left(  -t\right)
 e^{-i\omega_{k}t}\\
 &  = &  
n \left(  \omega_{k}\right)
 \frac{i}
{\omega - \omega_{k}+i\epsilon} -  \left( 1 + 
n \left(  \omega_{k}\right) \right)
 \frac{i}
{\omega-\omega_{k} - i\epsilon}.
\end{eqnarray}
For the other remaining Green functions, we have for positive 
$\omega$
\begin{equation}
G_{12}^{k}\left(  \omega\right)  = 2\pi n\left(  \omega_{k}\right)  
\delta\left(\omega-\omega_{k}\right),
\end{equation}
and
\begin{equation}
G_{21}^{k}\left(  \omega\right)  = 2\pi  \left( 1 +  
n\left(\omega_{k}\right)
\right)  \delta\left(
\omega-\omega_{k}\right).
\end{equation}
These Green functions are not all independent.  \ We first 
observe that  
\begin{equation}
G_{11}^{k}\left(  \omega\right)  + G_{22}^{k}\left(  \omega\right)  
= G_{12}^{k}\left(
\omega\right)  + G_{21}^{k}\left(  \omega\right).
\end{equation}
This is an immediate result of their 
definition. \ Moreover, the term 
on the l.h.s. is easily seen to be a symmetric sum of the product of two 
operators. \ Now it is not difficult to show 
from the above expressions 
of the Green functions that we have
\begin{equation}\label{fluct-diss-gen}
G_{11}^{k}\left(  \omega\right)  + G_{22}^{k}\left(  \omega\right) =
\left( 1+2n(\omega)  \right) \left[ G_{21}^{k}\left(  \omega\right)  
-  G_{12}^{k}\left(  \omega\right) \right].
\end{equation}
The last factor on the r.h.s. is an anti-symmetric 
sum of two operators. \ Equation $(\ref{fluct-diss-gen})$ is a 
statement of some form of the 
fluctuation dissipation theorem. \cite{martin}. 
\ These  
relations will be used in subsequent sections to calculate 
the correlation functions. \
In equilibrium, when
the distribution functions are the Bose-Enstein functions
\begin{equation}
1+2n(\omega)=\coth\frac{\beta\omega}2
\end{equation}
and equation $(\ref{fluct-diss-gen})$
gives the usual from of the fluctuation-dissipation theorem.
\bigskip

{\section{ Normal Mode Analysis }}

\bigskip

In this section we follow  Lyberatos, Berkov and Chantrell 
 \cite{lyberatos} and couple the normal modes of the spin 
system (also called collective field) 
to the harmonic oscillators of the bath. \ 
\ This method has also been recently used by Safonov and Bertram 
to calculate correlation functions of the magnetization 
in thin films. \cite{safonov} \ In this section, we show how 
their correlation functions for the collective 
field follow from the 
microscopic model treated here. \ The results in this section will be 
used in the next section to find the correlation functions of the 
magnetization in the general case.

\ First we find the collective 
degrees of freedom $c$ and $c^{\dagger}$ from 
the magnetization:
\begin{eqnarray}\label{Bogo}
a & = & u c + v c^{\dagger}  \label{transf} \\
a^{\dagger} & = & v c + u c^{\dagger}     \nonumber
\end{eqnarray}
where $u$ and $v$ are real. \ We require that 
\begin{equation}
\left[  c,c^{\dagger}\right]  =1.
\end{equation}
This implies that
\begin{equation}
u^{2}-v^{2} = 1.
\end{equation}
Therefore, we can write for some $\theta$,
\begin{eqnarray}
u &   = &  \cosh\theta  \\
v & =  & \sinh\theta  \nonumber.
\end{eqnarray}
If we set,
\begin{equation}
\omega_{0}^{2}=\Omega^{2}-\frac{K^{2}}{4},
\end{equation}
we find that the coefficients of the transformation Eq.$\left( 
\ref{transf} 
\right)$,
\begin{eqnarray}
u & = & \sqrt{\frac{\Omega+\omega_{0}}{2\omega_{0}}} ,\\
v & = & -\sqrt{\frac{\Omega-\omega_{0}}{2\omega_{0}}} \nonumber.
\end{eqnarray}
In this collective coordinates, the spin Hamiltonian becomes diagonal,
\begin{equation}
\mathcal{H}_{S}=\omega_{0} c^{\dagger}c.
\end{equation}
Now  the interaction term between the bath and the spin is 
taken of the form
\begin{equation}
\widehat{\mathcal{H}}_{SR} = - \sum_{k} \left( \gamma_{k} c^{\dagger}b_{k} + 
\gamma_{k}^{\ast}b_{k}^{\dagger}c \right).
\end{equation}
The generating functional for this system is then calculated in
Fourier space with the help of the Green functions stated in the 
last section. \ We find that 
\begin{eqnarray}
\widehat{\mathbb{Z}} \left[ \, \overline{\mathbf{J}}, \mathbf{J}\right] &  
= &   \int d\mu 
\left( \overline{z}_{1}, z_{1}  \right)\int
d\mu \left( \overline{z}_{2}, z_{2} \right)
 \exp\left\{  -\int\frac{d\omega}{2\pi}\; \left[ \;
\overline{\mathbf{Z}}\cdot \mathcal{A}\cdot
\mathbf{Z} + \overline{\mathbf{J}}\cdot \mathbf{Z} + \mathbf{J}\cdot
\overline{\mathbf{Z}} \; \right] \right\} \nonumber  \\
&  = & \frac{1}{\det \mathcal{A}}\exp\left\{- \int \frac{d\omega}{2 \pi} 
\; 
\overline{\mathbf{J}} 
\cdot \mathcal{A}^{-1} \cdot\mathbf{J}  \right\}
\end{eqnarray}

where  $\mathbf{Z}=\left( z_{1}, z_{2} \right)$ and $\mathbf{J}= 
\left( J_{1}, J_{2} \right) $ are  two-component vectors.
The matrix $\mathcal{A}$ is 
\begin{equation}
\mathcal{A}_{ij}=\left[
\begin{array}
[c]{cc}
i\left(  \omega_{0}-\omega\right)  +\Pi_{11}\left(  \omega\right)  
& \Pi_{12}\left(  \omega\right)\\
\Pi_{21}\left(  \omega\right) & - i \left(  \omega_{0}-\omega\right)  
+\Pi_{22}\left(  \omega\right)
\end{array}
\right].
\end{equation}
The $\Pi_{ij}$ terms are due to the interaction of the system with the 
bath.
\hspace{0pt}They depend on the density of states of the bath $\lambda 
(\omega)$ and the 
coupling
constants. \ For a general bath, the $Pi_{11}$ element is given   
\begin{eqnarray}
\label{Pi11}
\Pi_{11}\left(  \omega\right)   & = & \sum_{k}\left|  \gamma
_{k}\right|  ^{2}G_{11}^{k}\left(  \omega \right)  \\
& = &  i \int\frac{d\omega_{k}}{\pi}\pi\lambda\left(  \omega
_{k}\right)  \left|  \gamma\left( \omega_{k}\right)  \right|  ^{2}
 \left[  \frac{1 + n\left(  \omega_{k} \right)  
}{\omega-\omega_{k}+i\epsilon
}-\frac{n\left( \omega_{k}\right)  }{\omega-\omega_{k}-i\epsilon}\right].
\end{eqnarray}
The remaining matrix elements are calculated similarly. 
If now, we assume that the bath parameters satisfy the condition

\begin{equation}
\pi\lambda\left(  \omega_{k}\right)  \left|  \gamma \left( 
{\omega_{k}}\right)\right|
^{2}=\alpha  \;\;\;\;\;\; \left(  \omega_{k} > 0 \right) ,
\end{equation}
where $\alpha$ is a constant, we find that the 
interaction with the 
bath induces the following coefficients,
\begin{eqnarray}
\Pi_{11}\left(  \omega\right)&  = & \, \alpha
\left(1+2n (\omega)  \right)\\
\Pi_{22}\left(  \omega\right)  & = & \Pi_{11}\left(  \omega\right) \\
\Pi_{12}\left(  \omega\right)  & = & 2  \alpha n\left(  \omega\right) \\
\Pi_{21}\left(  \omega\right)  & = & 2  \alpha \left( 1 +
 n\left( \omega\right)\right) .
\end{eqnarray}
The correlation functions of the magnetization are related to the inverse 
elements of the matrix $\mathcal{A}$. \ A calculation of the inverse 
matrix gives

\begin{eqnarray}
\mathcal{A}^{-1}=\frac{1}{\mathcal{D}}\left[
\begin{array}
[c]{cc}
i \left(  \omega_{0}-\omega\right)  + \alpha  \left( 
1+2n(\omega) \right)    & -2  \alpha n\left(  \omega\right)  \\
- 2 \alpha \left( 1 + n\left( \omega\right) \right)   & - i 
\left(  \omega_{0} - 
\omega\right)  + \alpha\left( 1+2n(\omega) \right)
\end{array}
\right].
\end{eqnarray}
where $\mathcal{D}$ is the  determinant 
\begin{equation}
\mathcal{D}=\det \mathcal{A} = \left(  \omega_{0} - 
\omega\right)  ^{2}+\alpha^{2}.
\end{equation}

After solving for the generating functional in terms of the external 
sources,
we can expand it around the point $\mathbf{J}
=\mathbf{J}^{\ast}=0$ to get all the required correlation functions. \ 
In this case, we can solve 
for $\widehat{\mathbb{Z}}\left[ \, \overline{\mathbf{J}}, 
\mathbf{J}\right]$ 
exactly since the full 
Hamiltonian is quadratic. \  The correlation 
functions are now found by differentiations 
with respect 
to $ \overline{\mathbf{J}}$ and/or $\mathbf{J}$.
For $t > t^{\prime} $, we have for the correlation 
function of the collective operator
$c$

\begin{equation}
\left\langle c^{\dagger}(t)c(t^{\prime})\right\rangle =i  
\frac{\partial}{\partial
J_{2}(t)}\frac{\partial}{\partial\overline{J}_{1}(t^{\prime})}\left.  
\mathbb{W} \left[ \;
\overline{\mathbf{J}},\mathbf{J} \right]  \right|  
_{\mathbf{J}=\overline{\mathbf{J} }=0}
\end{equation}
where

\begin{equation}
\mathbb{W} \left[ \,  \overline{\mathbf{J}},\mathbf{J}\right]  = i 
\int\frac{d\omega}{2\pi}\, \overline{J}
_{i}\left(  \omega\right)  \mathcal{A}_{ij}^{-1}\left(  \omega\right)  
J_{j}\left(
\omega\right)
\end{equation}
All correlation functions of three operators or more are 
zero since the Hamiltonian is quadratic. \ The
above correlation function is therefore 
related to the matrix element,
$ \mathcal{A}^{-1}_{12}$. \ At high 
temperature, i.e., $\beta \rightarrow0\left(  \omega << kT\right) $
\begin{eqnarray}
\mathcal{A}_{12}^{-1}(\omega)  & = & - \frac{2\alpha n\left( 
\omega\right)  }{\left(
\omega_{0} - \omega\right)  ^{2}+\alpha^{2}}\\
& = &\frac{-2\alpha kT}{\omega}\frac{1}{\left(  \omega_{0} 
- \omega\right)
^{2}+\alpha^{2}}
\end{eqnarray}
In real time, we have 
\begin{eqnarray}
\mathbb{W}\left[ \,  \overline{J},J\right]  = i  
\int {d t}\int d t^{\prime} \, \overline{J}
_{i}\left( t \right)  \mathcal{A}_{ij}^{-1}\left( t - t^{\prime} 
\right)  J_{j}\left(
 t^{\prime} \right),
\end{eqnarray}
hence, the two-point correlation function of the field  $c$ associated 
with the $1-2$ component of the path $C$ is 
\begin{equation}
\left\langle c^{\dagger}(t) c(t^{\prime})  \right\rangle = 2 \alpha  
\int \frac{d \omega}{ 2 \pi } e^{-i \omega \left( t - t^{\prime} \right)}
\frac{n \left( \omega \right)}{\left(  \omega_{0}-\omega\right)
^{2}+\alpha^{2}}. 
\end{equation}
Similarly, we can find the corresponding $2-1$ component of the 
correlation function,
\begin{equation}
\left\langle  c(t^{\prime}) c^{\dagger}(t)\right\rangle = 2 \alpha  
\int \frac{d \omega}{ 2 \pi } e^{-i \omega \left( t - t^{\prime} \right)}
\frac{1 + n \left( \omega \right)}{\left(  \omega_{0}-\omega\right)
^{2}+\alpha^{2}}. 
\end{equation}
From these last two correlation function, we get the classical 
correlation function of $c$,
\begin{equation}\label{cc}
\frac{1}{2} \left\langle  c^{\dagger}(t) c(t^{\prime}) +   c(t^{\prime}) 
c^{\dagger}(t)  \right\rangle =  \alpha 
\int \frac{d \omega}{ 2 \pi } e^{-i \omega \left( t - t^{\prime} \right)}
\frac{1 + 2 n \left( \omega \right)}{\left(  \omega_{0}-\omega\right)
^{2}+\alpha^{2}}. 
\end{equation}
Now we observe that if we take 
the limits $\alpha \longrightarrow 0$ and 
$t^{\prime}  \longrightarrow t$, we recover the expectation
value of the occupation number $\widehat{n}=c c^{\dagger}$,
\begin{equation}
\langle \widehat{n} \left( \omega \right) \rangle = 1 + 
n\left( \omega \right).
\end{equation}
To get this limit, we have used the fact that 
\begin{equation*}
\frac{1}{\pi} \frac{\alpha}{\omega^{2}+\alpha^{2}} \longrightarrow
 \delta \left( \omega \right) \;\;\;\; \text{as}\;\;\; \alpha 
\longrightarrow 0.
\end{equation*}
We also note that in this limit, we  have
\begin{equation}
\langle c^{\dag}c + c c^{\dag} \rangle = 1 + 2 n \left( \omega \right),
\end{equation}
and 
\begin{equation}
\langle \lbrack c ( t ), c^{\dag}( t ) \rbrack \rangle = 1.
\end{equation}
This shows that the commutations relations and the FDT
are satisfied at all times $t$.

Since we are close to equilibrium and $\alpha$ is small, the power 
spectrum 
will be picked near $\omega=\omega_{0}$. \ Therefore, this 
is also equivalent to having a Langevin equation with random forces F such 
that \cite{haken}
\begin{equation}
\overset{.}{c}=-(i\omega_{0}+\alpha)c+F\left(  t\right)
\end{equation}
and
\begin{equation}
\left\langle \left\{  F^{\dagger}\left(  t\right)  ,F\left(  
t^{\prime}\right)
\right\}  \right\rangle = 2 \alpha k T \delta\left(  t-t^{\prime}\right).
\end{equation}
Within this approximation,  we recover the 
correlation functions of 
Safonov-Bertram, Eq.$\left( 2.14  \right)$ 
in Ref. \cite{safonov}. 

\begin{equation}
\left\langle c^{\dagger}(t)c(0)\right\rangle = \int 
\frac{d\omega}{2 \pi} 
\frac{2\alpha k T}{\omega_{0}} \frac{e^{-i \omega t }}{\left( 
 \omega_{0}-\omega \right)
^{2}+\alpha^{2} }.
\end{equation}

To get the 
correlation functions of the magnetization $\mathbf{S}$, 
we first use the linear 
transformation, Eq. $\left( \ref{transf} \right)$, to 
write $S_{x}$ and $S_{y}$ in terms of the collective operators 
$c$ and $c^{\dag}$. \ This result will be deduced in the next section
from the exact 
treatment of  the general asymmetric case and without 
recourse to the rotating wave approximation as was done in 
Ref. \cite{safonov}.

\bigskip

{\section{White and Colored Noise: LLG and Other Solutions}

\bigskip

One could compute the correlation functions for the original variables
from the $c c^\dag$ correlation functions, by inverting the Bogoliubov
transformation (\ref{Bogo}). However, in this section, we will repeat the
computation directly in terms of original magnetization component 
variables,
\begin{eqnarray}
\widehat{S}_{x} & = & \frac{1}{2}\left(  a + a^{\dagger}\right), \\
\widehat{S}_{y} & = & \frac{1}{2i}\left(  a - a^{\dagger}\right)\;.
\end{eqnarray}
Writing the generating functional $\overline{\mathbb{Z}}$ in terms 
of them is trivial, but performing the Gaussian integration is 
more complicated, since we will have to invert a $4\times4$ matrix. 
\ We couple the magnetization 
to an external 
time dependent magnetic field $\mathbf{F}$
\begin{equation}
\widehat{\mathcal{H}}_{e} = - \mathbf{F}\cdot 
\widehat{\mathbf{S}}.
\end{equation}
As we have seen in the previous section, in the 
generating functional approach, we double 
the components of $\mathbf{S}$ and that implies doubling 
of the external field $\mathbf{F}$. \ Therefore the 
interaction term becomes
\begin{equation}
\widehat{H}_{e}= -\mathbf{F}_{1}\cdot \mathbf{S}_{1} 
 + \mathbf{F}_{2}\cdot \mathbf{S}_{2}. 
\end{equation}   
The results for this Hamiltonian can be derived from those 
already found in the previous section.
The bath contributes a term of the following form to the
effective action

\begin{eqnarray}
S_{eff} \left( \mathbf{S}_{1}, \mathbf{S}_{2} \right) & = &
\left.\int\frac{d\omega}{2\pi}\left|  \gamma_{k}\right|  ^{2}
G_{11}^{k}\left( \omega \right)  \left[  S_{1,x}^{2} + S_{1,y}^{2}
+ i \overline{S}_{1,x}S_{1,y}-i \overline{S}_{1,y}S_{1,x}\right] \right.
  \\
&& -\int\frac{d\omega}{2\pi}\left|  \gamma_{k}\right|  ^{2}G_{22}^{k}\left(
\omega\right)  \left[  S_{2,x}^{2}+S_{2,y}^{2}
+  i \overline{S}_{2,x}S_{2,y} - i \overline{S}_{2,y}S_{2,x}\right]
 \nonumber \\
&& +\int\frac{d\omega}{2\pi}\left|  \gamma_{k}\right|  ^{2}G_{21}^{k}\left(
\omega\right)  \left[  \overline{S}_{2,x}S_{1,x} + \overline{S}_{2,y}
S_{1,y}
 + i \overline{S}_{2,x}S_{1,y} - i \overline{S}_{2,y}S_{1,x}\right]
 \nonumber \\
&& +\left. \int\frac{d\omega}{2\pi}\left|  \gamma_{k}\right| 
^{2}G_{12}^{k}\left(
\omega\right)  \left[  \overline{S}_{1,x}S_{2,x} + \overline{S}_{1,y}
S_{2,y}
 +i  \overline{S}_{1,x}S_{2,y} - i \overline{S}_{1,y}S_{2,x}\right]
   \right\}  \nonumber
\end{eqnarray}
where the bar denotes the complex conjugate integration variables and
$S_{1}$ $\left(  S_{2}\right)$ is the component along the path
$C^{(+)}\left(  C^{(-)}\right)$, Fig. \ref{path}. 
Next, we define two new vectors $\mathbf{S}$ and $\mathbf{D}$,
\begin{eqnarray}
\mathbf{S} & = & \frac{1}{2} \left( \mathbf{S}_{1} + \mathbf{S}_{2} 
\right),\\
\mathbf{D} & = &  \mathbf{S}_{1} - \mathbf{S}_{2}.
\end{eqnarray}
\noindent Similarly, we define
\begin{eqnarray}
\mathbf{F}_{d}  & = & \mathbf{F}_{1} - \mathbf{F}_{2} ,\\
\mathbf{F}_{a}  & = & \frac{1}{2}\left(  \mathbf{F}_{1} + 
\mathbf{F}_{2}\right).
\end{eqnarray}
Finally, we make another definition.\ We define four-vectors 
$\mathbf{U}$ and $\mathbf{F}$
\begin{eqnarray}
\mathbf{U}  & = & \left(  S_{x},S_{y}, D_{x},D_{y} \right), \\
\mathbf{F}  & = & \left(  \mathbf{F}_{d}, \mathbf{F}_{a} \right),
\end{eqnarray}
and write the generating functional in terms of these 
four-vectors along the path $C^{\left( + \right)}$ only. \ Since 
$\mathbf{U}\left( t \right)$ is real, then we have 
\begin{equation*}
\overline{U}\left( \omega \right) = U \left(- \omega \right),
\end{equation*}
and hence we should constrain the fourier integration to positive 
frequencies only. \
The bath-independent part of the 
Hamiltonian then gives the following contribution to the 
phase of $\widehat{\mathbb{Z}}$,
\begin{equation}
iI_{1}-iI_{2}=-  \int_{0}^{\infty}\frac{d\omega}{\pi}\overline{U}_{i}\left(
\omega\right)  \mathcal{A}_{ij}^{(0)}\left(  \omega\right)  
U_{j}\left(  \omega\right),
\end{equation}
where the matrix $\mathcal{A}^{(0)}$ is, in Fourier space, 
\begin{equation}
\mathcal{A}_{ij}^{(0)}=\left[
\begin{array}
[c]{cccc}%
0 & 0 & i A & -\omega\\
0 & 0 & \omega & i B\\
i A & -\omega & 0 & 0\\
\omega & i B & 0 & 0
\end{array}
\right].
\end{equation}
Again, it is the inverse of the full matrix $\mathbf{\mathcal{A}}
= \mathbf{\mathcal{A}}^{(0)} + \mathbf{\mathcal{A}}^{\text{int}}$,
that is needed to determine the correlation functions 
of the magnetization where \ $\mathbf{\mathcal{A}}^{\text{int}}$ is 
the part that is due to the interaction with the bath.
The determinant of  $\mathbf{\mathcal{A}}$ determines 
the natural frequency of the system and any broadening due 
to interactions. \ The determinant of the free part is 
\begin{equation}
\mathcal{D}_{0}=\left(  \omega_{0}^{2}-\omega^{2}\right)  ^{2},
\end{equation}
where
\begin{equation}
\omega_{0}^{2} = A B,
\end{equation} 
is the ferro-magnetic resonance (FMR) frequency of the system. \
The calculation of the matrix $\mathbf{\mathcal{A}}$ is 
done along the same lines as in the normal mode solution.

To recover dissipative behavior in the spin sub-system, we take 
the continuum limit in the number of oscillator modes. \ This 
limit guarantees that the probability of acquiring back 
any energy lost to the bath is zero. \ Because of the interaction
with the bath, we expect that there will be a shift in the 
energy of the spin system accompanied by dissipation. \

An explicit computation shows that in the continuum limit, i.e. converting 
the sum over $k$ in an integral over the frequencies involving 
the density of states $\lambda(\omega_k)$
\begin{equation}
\lambda(\omega_k)= \frac{dk}{d\omega_k} \;,
\end{equation}
the correlation functions can be expressed in terms of the 
functions $L_{r}$ and $L_{i}$ :
\begin{eqnarray}
L_{r} \left( \omega \right)  & = &  -i\int_0^\infty
\frac{d\omega_{k}}{\pi} \; \pi\lambda\left(  \omega_{k}\right)
\left|  \gamma\left(  \omega_{k}\right)  \right|  ^{2}(G_{11}^{k}-G_{22}^k)
\left(\omega_k\right), \\
L_{i} \left( \omega \right)  & = & 2\int_0^\infty 
\frac{d\omega_{k}}{\pi} \; \pi \lambda\left(  \omega_{k}\right)  \left|
\gamma\left(  \omega_{k}\right)  \right|  ^{2}(G_{12}^k-G_{21}^k)
\left(  \omega_k\right) .
\end{eqnarray}
Using the definitions of the Green functions, we find
\begin{eqnarray}
L_{r}\left(  \omega\right)   & = & 2\mathcal{P}
\int_0^\infty\frac{d\omega_{k}}{\pi}%
\pi\lambda\left(  \omega_{k}\right)  \left|  \gamma\left(  
\omega_{k}\right)
\right|  ^{2}\frac{1}{\omega-\omega_{k}}\;, \\
L_{i}\left(  \omega\right)   & = & -2\pi\lambda
\left(  \omega\right)  \left|
\gamma\left(  \omega\right)  \right|  ^{2}\theta(\omega)\;. \label{Li}
\end{eqnarray}
Counter-terms are needed to cancel ultraviolet divergences in
$L_{r}$. \ For simplicity, 
we will assume that this is done via suitable subtractions. The effect of 
$L_r$ is a redefinition of the given coefficients $A$ and $B$. \
This redefinition in principle changes the oscillation frequency. 
However, for a passive path, one 
neglects $L_r(\omega)$ and thus the frequency shift. In this approximation
 the coefficients  $A$ and  $B$ are kept unnormalized
   and all the physics is contained in 
$L_i(\omega)$. 
There is a subtlety here, since the expression (\ref{Li})
is not antisymmetric, whereas it has to be antisymmetric due to
general properties of correlation functions (see appendix for a
discussion). Therefore $L_i(\omega)$ has to be antisymetrized. By noticing
that $|\gamma(\omega)|^2$ is even in $\omega$ and extending 
$\lambda(\omega)$ to negative $\omega<0$ with a negative sign, the
final result can be written in the form
\begin{equation}
\mathcal{A}=\left[
\begin{array}
[c]{cccc}%
0 & 0 & i A - \Delta(\omega) &  - \omega\\
0 & 0 & \omega &  i B - \Delta(\omega)\\
i A + \Delta(\omega) & -\omega  & \pi\lambda\left(  \omega\right)  
\left|  \gamma\left(
\omega\right)  \right|  ^{2}\left(  1 + 2 n 
\left(  \omega\right)  \right)   & 0\\
 \omega & i B + \Delta(\omega)  & 0 & \pi\lambda\left(  \omega\right)  
\left|  
\gamma\left(
\omega\right)  \right|  ^{2}\left(  1 + 2 n\left(  \omega\right)  \right)
\end{array}
\right], \label{matrix}
\end{equation}
where $\Delta(\omega)$ is the odd function

\begin{eqnarray}
\Delta \left( \omega \right) &=& \frac{- L_{i} \left( \omega \right) +
  L_{i} \left( - \omega \right)}2=\pi\lambda(\omega)|\gamma(\omega)|^2\;.
\end{eqnarray}

\noindent The determinant of this matrix is given by
\begin{eqnarray}
\det \mathcal{A}=\mathcal{D}\left( \omega \right) & = & \left[  
\omega_{0}^{2}- \omega^{2} - \Delta \left(\omega \right)^{2} \right]^{2}
+ \left[ \Delta \left(\omega \right) \left( A + B \right)  
 \right]^{2}, \label{den}
\end{eqnarray}
 \ We also observe that for 
the functional integral to converge, we must have 
\begin{equation*}
\pi\lambda\left(  \omega\right)  \left|
\gamma\left(  \omega\right)  \right|  ^{2}\left(  1+2 n (\omega) \right)  
> 0.
\end{equation*} 
This requires that  the function  $L_{i}\left( \omega \right)$ when 
extended to negative 
frequencies be an {\it{odd}} function which is consistent 
with the statement 
before Eq.$(\ref{matrix})$. 

To calculate the correlation functions, we need  
first to calculate the inverse matrix of 
$\mathcal{A}$. \ The cofactors needed 
for the correlation functions
of the different components of the magnetization are for small 
couplings to the bath
\begin{eqnarray}
c_{11}& = & -\pi\lambda\left(  \omega\right)  \left|  
\gamma\left(  \omega\right)
\right|  ^{2}\left(  1+2 n ( {\omega}) \right)  
\left[  B^{2}
+\omega^{2}+ \Delta\left(\omega\right)^{2}  \right],\label{co11} \\
c_{12}  & = & i \pi\lambda\left(  \omega\right)  \left|  \gamma\left(
\omega\right)  \right|^{2} \left(  1+2 n ({\omega}) \right) 
 \left[   \left(  A+B \right) \omega \right], \\
c_{22} & = & - \pi\lambda\left(  \omega\right)  \left|  \gamma 
\left(  \omega\right)
\right|^{2}\left(  1+2 n ({\omega}) \right)   \left[  A^{2}
+\omega^{2} + \Delta \left(\omega\right)^{2} \right] ,\\
c_{12}  & = & - c_{21} = i \pi\lambda\left(  \omega\right) 
 \left|  \gamma\left(
\omega\right)  \right|^{2} \left(  1+2 n ({\omega}) \right) 
 \left[   \left(  A+B \right) \omega \right].
\end{eqnarray}
As will be seen below, the $c_{11}\; \left(c_{22} \right)$ 
co-factor of the matrix $\mathcal{A}$ is 
associated with the $xx \; \left( yy \right)-$component 
of the magnetization while $c_{12}$ is related to the 
$xy-$component. \

\bigskip

{\subsection*{ Correlation Functions}}

\bigskip

Next, we use these cofactors to 
calculate the correlation functions of 
the magnetization.

For  a general operator $\mathcal{O}$, the 
average of the anti- 
commutator $\left\{ \mathcal{O}(t), \mathcal{O} (t^{\prime}) \right\}$ 
is found by differentiation of $\mathbb{W}\left[ 
\mathbf{F}_{a}, \mathbf{F}_{d} \right]$ with respect to $\mathbf{F}_{d}$,
\begin{eqnarray}
\frac{1}{2}\langle \mathcal{O}\left(t \right) 
\mathcal{O}\left( t^{\prime} \right) + \mathcal{O}
\left(t^{\prime} \right) 
\mathcal{O} \left( t  \right) \rangle & = & -i \frac{\delta^{2}\mathbb{W}
\left[ \mathbf{F}_{a}, \mathbf{F}_{d} \right]}{\delta \mathbf{F}_{d}\left( 
t \right)
\delta \mathbf{F}_{d}\left( t^{\prime} \right)}  
\end{eqnarray}
Applying this procedure to the components 
of the magnetization, we find that for the $x-$component 
\begin{equation}
\frac{1}{2} \langle S_{x}(t)S_{x}(0)+S_{x}(0)S_{x}(t) 
\rangle
=\int\frac{d\omega}{2\pi}\cos\omega t \; 
\frac{c_{11}\left(  \omega\right)
}{\mathcal{D}\left(  \omega\right)  }.
\end{equation}

\noindent From Eq. $(\ref{co11})$, we then obtain

\begin{eqnarray}
C_{xx}(t)  & = & \int\frac{d\omega}{2\pi}\cos(\omega t) 
\left[  1+2 n ({\omega}) \right]  \pi\lambda\left(  \omega\right)  \left|
\gamma\left(  \omega\right)  \right|^{2}\; 
\frac{\left(  \omega^{2}+B^{2}\right)  + \Delta^{2}}{\mathcal{D}\left(
\omega\right)  }.
\end{eqnarray}
Now, we show how for different choices of the function $L_{i}\left(
\omega \right)$, we can recover the LLG result and other 
oscillator-type correlation functions.

{\subsubsection{ Case 1 : \; LLG}}

If we assume that the bath is defined such that 
\begin{equation}
\pi\lambda\left(  \omega\right)  \left|  \gamma\left(  \omega\right)  
\right|
^{2}=\alpha\omega,
\end{equation}
i.e., $L_{i}\left(\omega \right)$ is odd. Then 
in the limit of high 
temperature, $ \beta\rightarrow 0$,  the correlation function  
for the $xx$-component 
takes the simple form 

\begin{eqnarray}
C_{xx}(t)  & = &2\alpha kT\int\frac{d\omega}{2\pi}  \cos( \omega t)  \;
\left[ \frac{ \left(1+\alpha^{2}\right)\omega^{2}+B^{2}}
{\left[ \left(1+\alpha^{2}\right) \omega^{2}-\omega_{0}^{2}\right]
^{2}+\left[    \alpha\omega  \left(  A+B\right)  \right]  ^{2}
}\right].
\end{eqnarray}
This is the result that coincides with that derived from LLG. 
\cite{smith,rebei} \ This case also 
corresponds to a white noise solution.\cite{rebei}  \ Moreover, 
we observe that the condition on the bath that gives LLG 
is similar to the one that gave the harmonic oscillator 
solution, Eq.$\left( \ref{harmonic} \right)$. \ In both cases 
the spectral density is linear in frequency.\cite{grabert}

{\subsubsection{ Case 2 : \; Coherent Oscillator  }}

This case is similar to the normal mode result. \ We 
call it coherent oscillator because 
this case gives correlation functions similar to  those 
of the collective operator $c$ in the normal 
mode analysis. \ Here we choose $L_{i}(\omega)$ such that 

\begin{equation}\label{approx}
\pi\lambda\left(  \omega\right)  
\left|  \gamma\left(  \omega\right)  \right|
^{2}=\alpha, \;\;\;\; \omega >  0.
\end{equation}
This is the  same choice 
as in the previous section. 
For $ \beta\rightarrow0$, we get the following expression 
for the $xx$-component of the correlation functions

\begin{eqnarray}
C_{xx}(t)   = 2 \alpha k  T \;  \int\frac{d\omega
}{2\pi}  \cos(\omega t) \;\left(\frac{1}{\omega}\right)
\left[ \frac{B^{2}+ \omega^{2} + \alpha^2}{\left[  
\omega^{2} - 
\omega_{0}^{2}\right]^{2}+ 2 \alpha^{2} \left( \omega^{2} + 
\omega_{0}^{2} \right) + \alpha^{4}}\right].
\end{eqnarray}
\bigskip

\noindent  The normal mode result is easily seen to follow by setting 
$A=B$ in the correlation functions and $ ( 1/\omega) $ by $ (1/\omega_{0})$ 
. This result has been obtained before 
by Safonov and Bertram.\cite{safonov}  However without  this latter 
 approximation, 
this 
model corresponds to a case of colored noise.\cite{rebei} \ As $ \omega 
\rightarrow 0$, the integral diverges. \ Therefore at small $\omega$, the 
approximation in Eq. $( \ref{approx})$ is not 
applicable: $L_i(\omega)-L_i(-\omega)$ 
cannot be a constant but, for consistency with antisymmetry and 
analyticity,
must vanish with $\omega$ at $\omega\to0$.

\bigskip

{\section{ Conclusion}}

Starting from simple quantum models which only differ in how
the spin couples to the bath, we have been able 
to derive variant
correlation functions for the magnetization close 
to equilibrium. \ We have limited ourselves only to 
linear-type couplings. \ Depending on the 
coupling and the density of states of the 
bath, we showed how to obtain  different 
types of correlation functions including 
the classical LLG result. \ First, we showed that 
the typical harmonic oscillator correlation functions 
are recovered  only if the $S_{x}$ component of the magnetization
is coupled to the bath oscillators. \ Next, we coupled the normal 
modes of the spin to those of the bath and this allowed us to get 
the correlation functions of the collective field which is the 
starting point of the work of Safonov and Bertram. \ We were also 
able to use this special coupling to derive a more general type of 
correlation functions without recourse 
to any approximations such as the 
rotating wave approximation. \ These correlation functions are for 
a general linear coupling between the bath and the magnetization 
depend on the coupling constant $\gamma$ and 
the density of states of the bath system, $\lambda (\omega)$. \ For 
the $S_{x}S_{x}$-correlation function we find,

\begin{eqnarray} \label{Cxxfinal}
C_{xx}(t)   =  \int\frac{d\omega}{2\pi}
\left[  1+2n(\omega) \right]  \Delta \left( \omega \right)
  \cos (\omega t) \left[ \frac{  \omega^{2} + B^{2}   + \Delta
\left(\omega \right)^{2}  }{ \left(  
\omega_{0}^{2}- \omega^{2} - \Delta \left(\omega \right)^{2} \right)^{2}
+ \left[ \Delta \left(\omega \right) \left( A + B \right)  
 \right]^{2} } \right] .
\end{eqnarray}
Similarly for the $S_{y}S_{y}$-correlation function, we have
\begin{eqnarray}\label{Cyyfinal}
C_{yy}(t)   =  \int\frac{d\omega}{2\pi}
\left[ 1+2n( \omega) \right]  \Delta \left( \omega \right)
 \cos (\omega t)  \left[ \frac{ \omega^{2} + A^{2}   + 
 \Delta \left(\omega \right)^{2} }{\left(  
\omega_{0}^{2}- \omega^{2} - \Delta \left(\omega \right)^{2} \right)^{2}
+ \left[ \Delta \left(\omega \right) \left( A + B \right)  
 \right]^{2}  }\right] , 
\end{eqnarray}
and finally for the $S_{x}S_{y}$-correlation function, the 
correlation function is
\begin{eqnarray}\label{Cxyfinal}
C_{xy}(t)   & = &  \int \frac{d\omega}{2\pi} 
\left[ 1+2n(\omega ) \right]  \Delta 
\left( \omega \right) \sin (\omega t)  
 \left[ \frac{\omega\left(
A+B\right)  }{ \left(  
\omega_{0}^{2}- \omega^{2} - \Delta \left(\omega \right)^{2} \right)^{2}
+ \left[ \Delta \left(\omega \right) \left( A + B \right)  
 \right]^{2} } \right],
\end{eqnarray}
where 
\begin{equation}
\Delta \left( \omega \right) =  
\pi\lambda\left(  \omega\right)  \left|  \gamma\left(
\omega\right)  \right|^{2}.
\end{equation}

\bigskip

\noindent  \ The LLG solution was obtained for a special type of 
density of states and 
coupling to the bath. \ The same condition was also obtained in 
Ref. \cite{rebei} where in addition we were able to 
show that this choice gives the 
white noise character in the stochastic formulation 
. \ The normal mode solutions are however in general with memory. \ The
assumption that damping is constant close to the FMR frequency 
makes the equations of motion Markovian.\cite{peier} \ The damping 
in the cases treated here is {\it{independent}} of the symmetries of 
the Hamiltonian spin system, the reason being that the dissipation kernel
only depends on the coupling with the bath and the bath properties, but
not on the spin Hamiltonian.
\ For couplings other than linear, the 
damping is expected to depend on the symmetries of the full 
Hamiltonian, but, again, not at the leading order in perturbation theory. 
\ The point is that for non-linear coupling, the effective Hamiltonian
and therefore the correlation functions have to be computed perturbatively
in terms of Feynman diagrams; in particular the spin propagator will enter
in higher order computations and since the spin propagator depends on
the symmetry of the spin Hamiltonian, which could be isotropic ($A=B$)
or not ($A\neq B$), we will have different results for $A= B$ and 
$A \neq B$. \  This
does not happen at leading order in the non-linear case, and does not
happen \textit{at all orders} in the linear case, where the exact result
is available.

\ One last word about the non-equilibrium machinery 
used here to derive the above correlation functions: this choice of method 
allows us to go beyond the equilibrium formulation, and in particular to
show that a generalized fluctuation-dissipation theorem (\ref{fluct-diss-gen})
holds true even if the distribution functions are not exactly the
Bose-Einstein ones. \ In principle, an analysis of this system when the 
distribution functions present strong differences from the thermal one
is also possible in the general formalism we discussed here. However,
such a strongly out of equilibrium analysis would require further study
and it is outside the scope of the present paper.

\section*{acknowledgments}

We thank D. Boyanovsky and W. Hitchon for stimulating discussions. \ Comments
made by A. Lyberatos are also appreciated.
One of the authors (A.R) would like to thank R. Chantrell for 
discussions and is grateful to L. Benkhemis for help at various stages 
of this work.



\section*{Appendix}

Here we briefly show how the correlation functions derived 
in the main text can be derived using the equilibrium imaginary
time formalism \cite{orland} and without going to the coherent 
state representation. This is an equilibrium consistency check for our 
non-equilibrium computation.

The basic idea in the equilibrium computation is to invoke the
fluctuation-dissipation theorem (which is \textit{not} assumed in the
non-equilibrium computation) in order to derive the fluctuations from
the dissipation, i.e. from the spectral density.
The fluctuation-dissipation theorem says that in equilibrium 
the symmetric correlation functions can be written in
terms of the spectral densities $\rho^{ij}(\omega)$ as
\begin{equation}
\label{fluct-diss}
<\{S^i(t),S^j(0)\}>=
\int d\omega\; e^{i\omega 
t}\;\coth\frac{\beta\omega}2\;\rho^{ij}(\omega)\;,
\end{equation}

\noindent where $i$ and $j$ denote the indices $x$ and $y$ respectively.
From this definition it is immediate to see that the spectral densities
must satisfy the relationships 
\begin{equation}
\rho^{ij}(\omega)^*=\rho^{ji}(\omega),\quad \rho^{ij}(\omega)=
-\rho^{ji}(-\omega) .
\end{equation}
In particular, $\rho^{xx}$ and $\rho^{yy}$ are 
real and antisymmetric:
\begin{equation}
\rho^{ii}(\omega)^*=\rho^{ii}(\omega),\quad \rho^{ii}(\omega)=
-\rho^{ii}(-\omega).
\end{equation}
Thus, one can extract the spectral densities $\rho^{ii}(\omega)$
from the spectral representation of the retarded self-energy,
\begin{equation}
D_R^{ij}(\omega)=\int d\omega'\;\frac{\rho^{ij}(\omega')}{\omega'-\omega+i
\varepsilon}
\end{equation}
by taking the imaginary part:
\begin{equation}
\rho^{ii}(\omega)=-\frac1\pi\mbox{ Im } D_R^{ii}(\omega)\;.
\end{equation}
Moreover, due to the behavior of the theory under time reflections 
$t\to-t$,
\begin{equation}
<\{S^x(t)\,,\,S^y(0)\}>=-<\{S^x(-t)\,,\,S^x(0)\}>
\end{equation}
we have that  $\rho^{xy}(\omega)$ and $\rho^{yx}(\omega)$ 
are imaginary and symmetric:
\begin{equation}
\rho^{ij}(\omega)^*=-\rho^{ij}(\omega),\quad \rho^{ij}(\omega)=
\rho^{ij}(-\omega),\quad i\neq j
\end{equation}
As a consequence, the $\rho^{ij}(\omega),\ i\neq j$ spectral densities can 
be
extracted from the \textit{real} part of the retarded self-energy:
\begin{equation}
\rho^{ij}(\omega)=\frac i\pi\mbox{ Re } D_R^{ij}(\omega),\quad i\neq j
\end{equation}
In order to compute the retarded propagators, one has to compute the 
Euclidean effective action obtained by integrating
out the bath degrees of freedom in the Euclidean functional integral
\begin{equation}
e^{-S^E_{eff}(S^i)}=\int[db_k^* db_k]\exp\left[-\int_0^\beta d\tau\; 
L^E(S^i,b_k,b_k^*)\right]
\end{equation}
where
\begin{equation}
L^E(S^i,b_k,b_k^*)=L^E_S(S^i)+L^E_R(b_k,b^*_k)+L^E_{SR}(S^i,b_k,b_k^*),
\end{equation}
with
\begin{eqnarray}
L^E_S(S^i)&=&S^x i\partial_\tau S^y+\frac12 A(S^x)^2+\frac12
B(S^y)^2 ,\\
L^E_R(b_k,b_k^*)&=&\sum_k b^*_k (\partial_\tau-\omega_k) b_k ,\\
L^E_{SR}(S^i,b_k,b_k^*)&=&\sum_k b_k^*\gamma_k S_- + S_+\gamma_k^* b_k .
\end{eqnarray}
Since the integration on $b_k$ and $b_k^*$ is Gaussian, 
$S_{eff}$ can be computed exactly and
is quadratic in the spin fields:
\begin{equation}
S^E_{eff}(S^i)=\int_0^\beta d\tau \int_0^\beta d\tau'\;\frac12\; 
S^i(\tau)\; D_{ij}^E(\tau-\tau')\;S^j(\tau') .
\end{equation}
The Euclidean propagator is easily obtained in the Matsubara formulation
by inverting the 2x2 matrix
\begin{equation}
D_E^{-1}(\omega_n)=\begin{pmatrix}A & \omega_n\cr -\omega_n 
&B\end{pmatrix}+
\begin{pmatrix}\Pi_E(\omega_n) & 0\cr 0 &\Pi_E(\omega_n)\end{pmatrix}
\end{equation}
where the first matrix is the inverse free propagator and the second
matrix is the self-energy matrix (to be compared with the real time
result Eq. (\ref{Pi11}))

\begin{equation}
\Pi_E(\omega_n)=2\sum_k|\gamma_k|^2 G_E^k(\omega_n)
\end{equation}
and where $\omega_n=2\pi n T,\ n=0,\pm1,\pm2,\dots$
are the Matsubara frequencies. The inversion is trivial. \
The retarted propagator can be obtained with an analytic continuation
$\omega_n\to i\omega$:
\begin{equation}
D_R(\omega)=\frac1{\mathcal{D}(\omega)}
\begin{pmatrix}B+\Pi_R(\omega) & -i\omega\cr i\omega & A+\Pi_R(\omega)
\end{pmatrix}
\end{equation}
where $\mathcal{D}(\omega)$ is the determinant
\begin{equation}
\mathcal{D}(\omega)=\omega_0^2-\omega^2+(A+B)\Pi_R(\omega)+\Pi_R^2(\omega)
\end{equation}
In order to compute real and imaginary parts, it is convenient to 
split
\begin{equation}
\Pi_R(\omega)=\mbox{ Re }\Pi(\omega)+i\mbox{ Im }\Pi(\omega)
\end{equation}
and to introduce the real quantities
\begin{equation}
\tilde A(\omega)=A+\mbox{ Re }\Pi(\omega),\quad \tilde B(\omega)=B+
\mbox{ Re }\Pi(\omega)
\end{equation}
Then the inverse determinant reads
\begin{equation}
\mathcal{D}^{-1}(\omega)=\frac1{\omega_0^2-\omega^2-|\Pi_R|^2+
(\tilde A+\tilde B)\Pi_R}=
\frac{\omega_0^2-\omega^2-|\Pi_R|^2+(\tilde A+\tilde B)
(\mbox{ Re }\Pi-i\mbox{ Im }\Pi)}
{|\mathcal{D}(\omega)|^{2}}
\end{equation}
with
\begin{equation}
|\mathcal{D}(\omega)|^2=
[\omega_0^2-\omega^2-|\Pi_R(\omega)|^2+\mbox{ Re }\Pi(\omega)
(\tilde A+\tilde B)]^2+
(\tilde A+\tilde B)^2(\mbox{Im }\Pi(\omega))^2
\end{equation}
The function $\mbox{ Re }\Pi(\omega)$ and $\mbox{ Im }\Pi(\omega)$ 
are related to the previously
defined functions $L_r(\omega)$, $L_i(\omega)$ and $\Delta(\omega)$. 
In particular
\begin{equation}
\mbox{ Im }\Pi(\omega)=-2\pi\sum_k|\gamma_k|^2\delta(\omega-\omega_k)=-
\pi\lambda(\omega)|\gamma(\omega)|^2=-\Delta(\omega)\;.
\end{equation}
Notice that the continuum limit has been taken by ensuring the antisymmetry
of $\mbox{ Im }\Pi(\omega)$.
The $\rho^{ii}(\omega)$ spectral densities are obtained by taking the 
imaginary part of the full retarded propagators $D_R^{ii}(\omega)$:

\begin{equation}
\rho^{xx}=-\frac{1}{\pi|\mathcal{D}|^2}\left[
\omega_0^2-\omega^2-|\Pi_R|^2-B(
\tilde{ A}+
\tilde{B})\right]\Delta
\end{equation}

\begin{equation}
\rho^{yy}=-\frac{1}{\pi|\mathcal{D}|^2}\left[
\omega_0^2-\omega^2-|\Pi_R|^2-A(
\tilde A+
\tilde B)\right]\Delta
\end{equation}
whereas the spectral densities, $\rho^{ij}(\omega)$ ($ i\neq j$),
 are obtained by taking the
real part of $D_R^{ii}(\omega)$:

\begin{equation}
\rho^{xy}(\omega)=\frac i{\pi|\mathcal{D}|^2}\left[ 
\omega(\tilde A+\tilde 
B\right] \Delta)
\end{equation}
These results coincide with equations 
(\ref{Cxxfinal}, \ref{Cyyfinal}, \ref{Cxyfinal}). 
Therefore there is
full consistency between the real time and the imaginary time formalism.

The LLG limit for small damping 
is recovered  
when $\mbox{ Re }\Pi\to const,\; \mbox{ Im }\Pi\to\alpha\;\omega$, 
and the coherent oscillator is recovered in the region 
$|\omega|\sim\omega_0$ 
when $\mbox{ Re }\Pi\to const,\; \mbox{ Im }\Pi\to\alpha\;\mbox{sgn} 
\omega$. \ The term  $\mbox{ Re }\Pi$ is usually set to zero after being 
absorbed in the definition of the FMR frequency.

\newpage

\begin{figure}[ptb]
\resizebox{\textwidth}{!}
{\includegraphics[0in,0in][8in,9in]{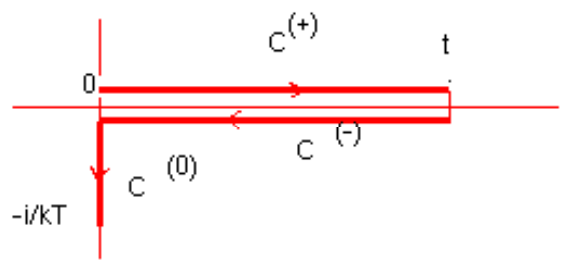}}
\caption{Complex time-path for the 
generating functional}
\label{path}
\end{figure}

\end{document}